\documentclass[aps,prl,reprint,groupedaddress]{revtex4-1}

\usepackage{microtype}
\usepackage{amsmath,amssymb,amsfonts,amsthm,mathtools}
\usepackage[colorlinks=true, linkcolor = blue, hyperfootnotes=false, citecolor = blue]{hyperref}
\usepackage[english]{babel}
\usepackage{graphicx}
\usepackage{bm}
\renewcommand\[{\begin{equation}}
\renewcommand\]{\end{equation}}
\renewcommand{\t}[1]{\mathbf{#1}}
\newcommand{\gt}[1]{\bm{#1}}

\begin{document}
\title{Polar metamaterials: A new outlook on resonance for cloaking applications}
\author{H. Nassar}\email{nassarh@missouri.edu}
\author{Y.Y. Chen} \email{yc896@missouri.edu}
\author{G.L. Huang}\email{huangg@missouri.edu}
\affiliation{Department of Mechanical and Aerospace Engineering, University of Missouri, Columbia, MO 65211, USA}
\begin{abstract}
Rotationally resonant metamaterials are leveraged to answer a longstanding question regarding the existence of transformation-invariant elastic materials and the ad-hoc possibility of transformation-based passive cloaking in full plane elastodynamics. Combined with tailored lattice geometries, rotational resonance is found to induce a polar and chiral behavior; that is a behavior lacking stress and mirror symmetries, respectively. The central, and simple, idea is that a population of rotating resonators can exert a density of body torques strong enough to modify the balance of angular momentum on which hang these symmetries. The obtained polar metamaterials are used as building blocks of a cloaking device. Numerical tests show satisfactory cloaking performance under pressure and shear probing waves, further coupled through a free boundary. The work sheds new light on the phenomenon of resonance in metamaterials and should help put transformation elastodynamics on equal footing with transformation acoustics and optics.
\end{abstract}
\maketitle
The peculiar behavior of locally resonant metamaterials has long been understood within Cauchy's model of elasticity albeit with constitutive properties that are frequency-dependent and possibly negative~\cite{Ma2016, Zhou2012}. Thus, for instance, when the elasticity (bulk or shear) modulus and mass density are both positive, the metamaterial exhibits a dispersive passing band; when either one is negative but not the other, the metamaterial exhibits a stop band; and when both are negative, the metamaterial exhibits a passing band with a negative refractive index~\cite{Liu2011, Zhu2014}. However, as we show here, resonance has other profound effects on wave motion that escape the paradigm of frequency-dependent effective parameters. In particular, we demonstrate how rotational resonance modifies not only the values of the effective moduli but the very form of the elasticity tensor as well.

The conventional form of the elasticity tensor is strongly constrained by a number of symmetries~\cite{landau1970}. The minor index symmetries ($a_{ijkl}=a_{jikl}=a_{ijlk}$) are equivalent to the symmetry of Cauchy's stress tensor; they stem from the balance of angular momentum and reduce the number of independent coefficients in a two-dimensional elasticity tensor $\t a = a_{ijkl}$ from $16$ to $9$. The major index symmetry ($a_{ijkl}=a_{klij}$) derives from the existence of a strain energy and further reduces that number to $6$. Then isotropy, understood as invariance under proper rotations, specifies the form of $\t a$ into
\[
a_{ijkl} = \mu(\delta_{ik}\delta_{jl}+\delta_{il}\delta_{jk}) + (\kappa-\mu)\delta_{ij}\delta_{kl}
\]
leaving only $2$ independent coefficients, the bulk and shear moduli $\kappa$ and $\mu$. Notably, without asking for it, $\t a$ has mirror symmetry~\cite{Forte1996, He1996}. Rotational resonance shakes this state of affairs: it generates torques strong enough to modify the balance of angular momentum thus dissolving stress symmetry and, with it, mirror symmetry. The elasticity tensor $\t a$ then gains two resonance-induced degrees of freedom, one quantifying chiral effects due to the loss of mirror symmetry and one quantifying polar effects due to the loss of stress symmetry.
\begin{figure}[t!]
\includegraphics[width=\linewidth]{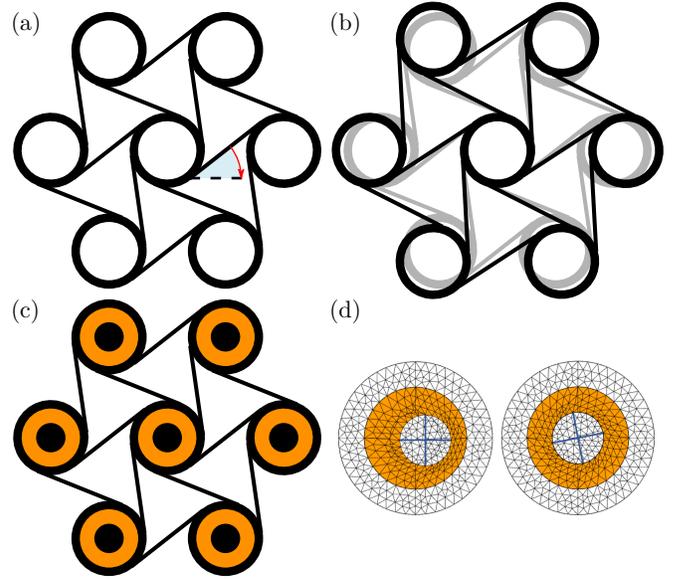}
\caption{Hexachiral lattices. (a) Nonresonant lattice: black circles are rigid masses and black segments are massless elastic rods; contact points are perfect hinges. The highlighted angle is the twisting angle $\gamma$. (b) Isotropic zero mode: the lattice expands (black) from its reference configuration (gray) without stretching any rods. (c) Polar metamaterial: black disks are embedded resonators; colored coating is elastic and massless. (d) Resonance modes under imposed outer displacement: translational ($\omega_t$) to the left and rotational ($\omega_r$) to the right.}
\label{fig:lattice}
\end{figure}

Accordingly, resonance effectively widens the design space of elastic materials to include polar metamaterials, i.e., metamaterials with dynamically broken stress symmetry. Within that space, we find long sought candidates for the realization of elastic cloaks in the form of hexachiral lattices with rotating inner resonators. We numerically demonstrate how, at the resonance frequency, a spatial gradient of these lattices can guide coupled pressure and shear waves around a cloaked region with minimal scattering. The theory underpinning the proposed design has been described in earlier work~\cite{GVasquez2012, Nassar2019} and substantiated leveraging grounded torsional springs for the production of the necessary torques~\cite{Nassar2018a}. Here, thanks to resonance, we bypass the need for a ground and provide a first portable solution.

Consider the chiral lattice sketched on Figure~\ref{fig:lattice}a for starters. Each of its unit cells hosts a rigid mass capable of translation and rotation in the plane and connected through hinges to a set of massless elastic rods. We are interested in the behavior of the lattice in the homogenization limit, i.e., when the unit cell size is infinitely small compared to the typical propagated wavelength. In that regime, the lattice displays an elastic behavior with elasticity tensor $\t a$ given by
\[
a_{ijkl} = \mu(\delta_{ik}\delta_{jl}+\delta_{il}\delta_{jk}-\delta_{ij}\delta_{kl})
\]
where
\[\label{eq:shear}
\mu = k\sqrt{3}/4
\]
is shear modulus, $k$ being the spring constant of the rods. The bulk modulus on the other hand is zero because the lattice can expand equally in all directions without stretching any rods (Figure~\ref{fig:lattice}b). More relevant to our purposes is the following conundrum: while the lattice is chiral for any twisting angle $\gamma\neq 0$, tensor $\t a$ is systematically mirror-symmetric. This is due to the conspiracy of space dimensionality, the index symmetries and isotropy, as portrayed in the introduction. Together, these constraints leave no room in $\t a$ to express a chiral behavior.

Next, let us embed within each mass a rigid core coated with a soft light material (Figure~\ref{fig:lattice}c). The obtained mass-in-mass resonator has two resonance frequencies $\omega_t$ and $\omega_r$ corresponding to translational and rotational oscillation modes and depicted on Figure~\ref{fig:lattice}d (see, e.g.,~\cite{Liu2005, Favier2018} and~\cite[App. A]{Note1} for expressions). The translational motion of the core modifies the balance of linear momentum. It generates a frequency-dependent body force which can be absorbed into the expression of effective mass density
\[\label{eq:MD}
\rho = \frac{M}{A} + \frac{m/A}{1-\omega^2/\omega_t^2},
\]
where $M$ and $m$ are the outer and inner masses, respectively, $A$ is the unit cell area and $\omega$ is angular frequency. Otherwise, the translational motion of the inner mass has no influence on $\t a$. By contrast, the rotational motion of the inner mass modifies the balance of angular momentum. As mass $M$ rotates through an angle $\Psi$, it perceives a restoring torque equal to $-\eta\Psi$ where
\[
\eta = -\omega^2\frac{i}{1-\omega^2/\omega_r^2}
\]
is an effective torsion stiffness and $i$ is the moment of inertia of the inner mass~\cite[App. B]{Note1}. It is important to remain consistent with the premises of homogenization here by maintaining that $\omega$ is infinitely small when compared to the typical cutoff frequency $\omega_c\propto\sqrt{k/M}$. Accordingly, the torques $-\eta\Psi\propto \omega^2i$ can be neglected except near the rotational resonance where the small denominator $1-\omega^2/\omega^2_r$ conveniently tames the small numerator $\omega^2i$. This is assumed henceforth; specifically, we work in the asymptotic regime $\omega=\omega_r+\delta\omega\ll\omega_c$ where
$\delta\omega/\omega_c\propto \omega_r^3/\omega_c^3$ and the torsional stiffness
\[
\eta = \frac{\omega_r^3}{2\delta\omega}i = \frac{\omega_r^3/\omega_c^3}{2\delta\omega/\omega_c}\omega_c^2i
\]
is of leading order. With that in mind, and by Cauchy's second law of motion, the stress tensor $\gt\sigma$ satisfies
\[
\sigma_{12}-\sigma_{21} = -\frac{\eta\Psi}{A},
\]
that is: the asymmetric part of stress is equal to the externally applied torque density. What is more is that $\t a$ then no longer needs to enforce the minor symmetries, meaning that, while remaining isotropic, it can in principle express chirality.
\begin{figure}[t!]
\includegraphics[width=\linewidth]{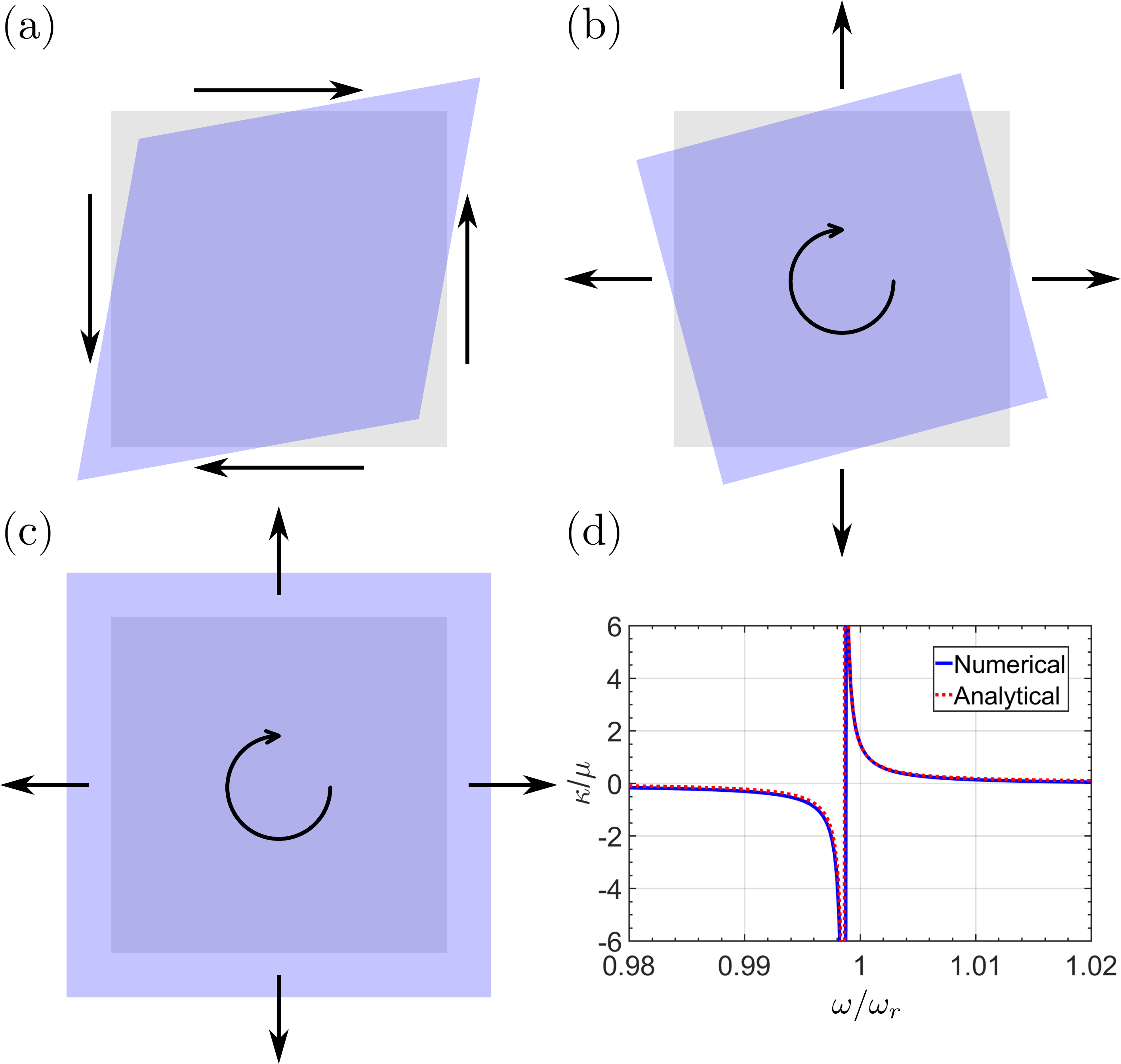}
\caption{Effective constitutive law~(\ref{eq:ECL}): response under (a) a pure shear, (b) a rigid rotation and (c) a pure dilation; deformations are illustrated on the geometry; arrows represent forces and torques. (d) Typical profile of the effective dynamic bulk modulus in the vicinity of $\omega_r$.}
\label{fig:behavior}
\end{figure}

Direct calculations confirm these predictions. We find
\[\label{eq:ElaTensor}
a_{ijkl} = \mu\left(\frac{\eta-12b^2k}{\eta+12b^2k}R_{ji}R_{lk}+R_{li}R_{jk}+\delta_{ik}\delta_{jl}\right),
\]
where $b$ is the radius of $M$ and $\t R$ is the plane rotation of angle $\gamma$~\cite[App. C]{Note1}. Since all plane rotations commute, it is straightforward to check that $\t a$ is isotropic. Furthermore,
\[
a_{1112} = -a_{1121} = \frac{2\eta\mu}{\eta+12b^2k}\sin\gamma\cos\gamma \neq 0
\]
shows at once that $\t a$ is both chiral and without minor symmetries so long as $\gamma\neq 0$. Rather than inspect $\t a$ however, it is far more insightful to write the stress-strain relationship, say decomposed into three parts, deviatoric, hydrostatic and skew, as in
\[\label{eq:ECL}
\tilde{\gt\sigma} = 2\mu\tilde{\t e},\quad
p = -\kappa t-\beta\theta,\quad
c = -\beta t - \alpha\theta;
\]
see Figure~\ref{fig:behavior}a-c. Above, ($i$) $\mu$ is a shear modulus that transforms a deviatoric strain $\tilde{\t e}$ into a deviatoric stress $\tilde{\gt\sigma}$; ($ii$) $\kappa$ is a bulk modulus that transforms an infinitesimal change in area $t$ into a hydrostatic pressure $p$; ($iii$) $\alpha$ is a polarity modulus that transforms an infinitesimal rotation $\theta$ into a body torque $c$; and ($iv$) $\beta$ is a chiral coupling that transforms a change in area into a body torque and an infinitesimal rotation into a hydrostatic pressure. While $\mu$ remains as it was, the other constitutive parameters have the expressions
\[
\kappa = \kappa_o\cos^2\gamma,\quad
\alpha = 4\kappa_o\sin^2\gamma,\quad
\beta = -2\kappa_o\cos\gamma\sin\gamma,
\]
with
\[\label{eq:bulk}
\kappa_o = \frac{\eta}{\eta+12b^2k}\frac{k\sqrt{3}}{2}.
\]

Variations on hexachiral lattices have been previously modeled as metamaterials with singly or doubly negative properties~\citep{Liu2011a, Liu2011, Zhu2014}, as micropolar media~\citep{Spadoni2012, Liu2012a, Chen2014b, Frenzel2017} or as strain-gradient media~\cite{Auffray2010, Bacigalupo2014, Rosi2016} based on different homogenization schemes. The present asymptotic analysis shows that the proposed resonant hexachiral lattices are best understood, in the immediate vicinity of rotational resonance, as polar metamaterials, i.e., metamaterials with broken stress symmetry. The analysis is validated numerically by loading a single unit cell and plotting its response against frequency. The results for $\kappa$ are shown on Figure~\ref{fig:behavior}d; the other parameters have identical profiles up to a scaling factor. It is noteworthy that near resonance, $\kappa$ changes sign suggesting that, for suitably chosen $\omega_t$ and $\omega_r$, a negative refraction band can be obtained. While this is true, it holds not because $\kappa$ becomes negative per se but because $\kappa_o$ does. Indeed, the phase velocities squared are equal to $\mu/\rho$ and $(\mu+\kappa_o)/\rho$ and do not involve $\kappa$ directly. More importantly, an analysis based on dispersion bands alone would give access to $\mu$, $\kappa_o$ and $\rho$, at best, but cannot reveal any of the interactions dictated by $\alpha$ and $\beta$ nor uncover the underlying broken symmetries. In particular, the fact that $\t a$ is isotropic and has broken minor symmetries makes the underlying lattice suitable for the design of elastic invisibility cloaks using conformal transformations~\cite{Nassar2019}. This is pursued next.

\begin{figure}[t!]
\includegraphics[width=\linewidth]{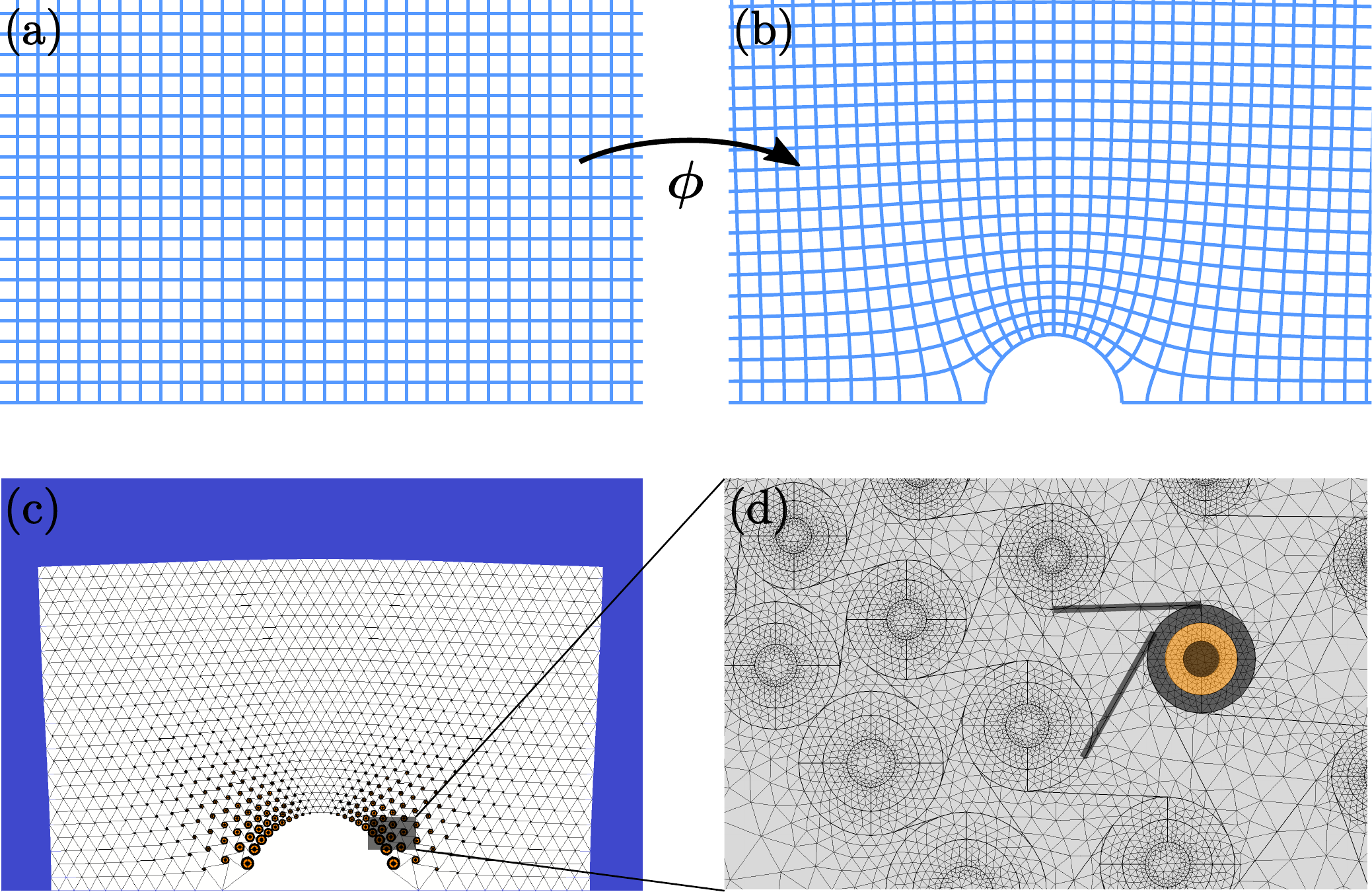}
\caption{Cloak's geometry. (a) Original domain: bottom side is free; other sides extend to infinity. Lines can be thought of as rays, coordinate axes or a discretization grid. (b) Image (transformed) domain: $\gt\phi$ opens a cave along the free boundary; rays are bent around the cave and back into their original paths as if the cave was not there. (c) Image domain with the cloak's microstructure resolved: the lattice is graded so as to fit the required properties at each position and is truncated beyond a given thickness. (d) Mesh used in the finite element analysis: a close up of the region framed in (c); a few structural elements are highlighted to facilitate their identification.}
\label{fig:mesh}
\end{figure}

A conformal transformation $\t x=\gt\phi(\t X)$ preserves angles~\cite{Norris2012, Xu2014}. Its gradient is therefore equal to a dilation of factor $\lambda$ composed with a rotation $\t R$. Starting with an original medium $\{\t X\}$ of shear modulus $\mu_o$, bulk modulus $\kappa_o$ and mass density $\rho_o$, composing a stretch
\[\label{eq:mass}
\lambda=\sqrt{\rho_o/\rho}
\]
with a rotation of angle $\gamma$, we retrieve an image medium $\{\t x\}$ satisfying equations~(\ref{eq:MD}) and~(\ref{eq:ECL}) exactly~\cite{Nassar2019}. Then, by the principles of transformation-based cloaking, domains $\{\t X\}$ and $\{\t x\}$ will be indistinguishable~\cite{Greenleaf2003a, Greenleaf2003, Leonhardt2006, Pendry2006}. For our demonstration, the original domain is a free half-space. The adopted transformation then creates a ``cave'' along the free boundary where things can hide without being detected by probing waves (Figure~\ref{fig:mesh}a,b). Clearly, the transformation is not uniform. As $\lambda=\lambda(\t x)$ and $\gamma=\gamma(\t x)$ change from one point $\t x$ to another, the underlying lattice will need to be graded in space (Figure~\ref{fig:mesh}c). This can be done once a small uniform discretization step $\delta$ of the original domain has been selected. Then, the lattice parameter near point $\t x$ is simply $\delta\lambda(\t x)$. This allows to calculate the outer radius $b(\t x)=\delta\lambda(\t x)\sin(\gamma(\t x))/2$ of mass $M$ and the unit cell area $A(\t x)=\delta^2\lambda^2(\t x)\sqrt{3}/2$. The inner radius of $M$ is arbitrarily set to $2b/3$ and the radius of the core is set to $b/3$ so that the coat's thickness is $b/3$ as well.
\begin{figure*}[t!]
\includegraphics[width=\textwidth]{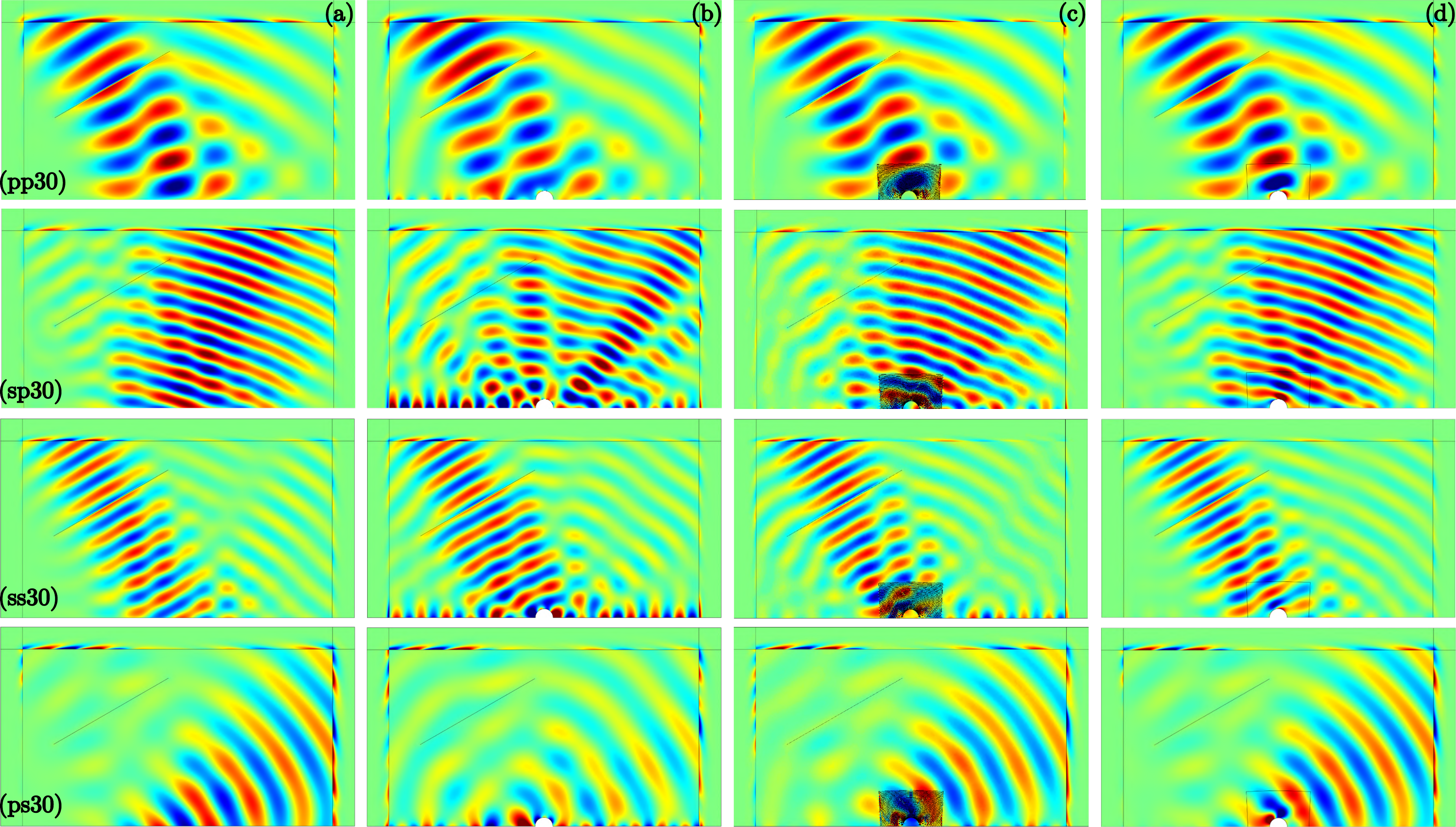}
\caption{Cloaking simulations under $30^\circ$ incidence: row (ij30), where i and j refer to pressure ``p'' or shear ``s'', depicts the component of type i under an incident wave of type j. Columns (a-d) correspond respectively to the original domain unaltered, the original domain with an uncloaked void, the void cloaked with a discrete lattice and the void cloaked with a fictitious continuous material. The numerical parameters are: $\omega_r=2\pi\times25\mathrm{kHz}$; $\rho_o=2.7\mathrm{kg/m^2}$; $\mu_o=25\mathrm{MPa.m}$; the cave is semi-circular with a $40\mathrm{mm}$ radius; the cloak is approximately $280\mathrm{mm}\times 155\mathrm{mm}$ and contains about $44\times26$ unit cells.}
\label{fig:clkSimP}
\end{figure*}

The above scheme determines unambiguously the geometry of the cloaking lattice. As for its constitutive parameters, they need to solve equations~(\ref{eq:shear}),~(\ref{eq:bulk}) and~(\ref{eq:mass}). For simplicity, we select a background medium such that $\kappa_o=2\mu_o$ in which case it is possible to work at the exact rotational resonance frequency $\omega=\omega_r$. Thus, $\eta$ is infinite and the spring constant of the lattice can be determined easily to be $k=2\kappa_o/\sqrt{3}$; in particular, it is uniform throughout the cloak. It is also possible to keep uniform the Lamé parameters of the core's coat; then, the inner mass $m=m(\t x)$ is graded so as to produce the same $\omega_r$ for all resonators. Last, $M=M(\t x)$ is calculated by solving for $M$ in~(\ref{eq:mass}).

We simulated the cloaking of the cave against probing pressure and shear waves at various angles of incidence using the finite element solver COMSOL Multiphysics. For convenience, the whole domain was meshed, including the background medium, the lattice, the empty space between the masses and the cloaked cave (Figure~\ref{fig:mesh}d). Empty spaces were assigned a massless compliant material whereas the rigid inner and outer masses were assigned a stiff material. The action of the elastic rods is taken into account in the weak form of the equations by integration along the straight lines in the mesh that connect the various neighboring masses. Besides the free boundary, perfectly matched layers are appended to the other three boundaries to suppress parasitic reflections.

The results are summarized on Figure~\ref{fig:clkSimP} for pressure and shear probing waves coupled to one another through the free boundary and incident at $30^\circ$; for other angles, see~\cite[App. D]{Note1}. While not perfect, the designed resonant lattice shows satisfactory cloaking performance as it suppresses pressure-shear (plot (sp30-b) vs (sp30-c)), shear-pressure (plot (ps30-b) vs (ps30-c)) and Rayleigh (plots (sp30-b) vs (sp30-c) and (ss30-b) vs (ss30-c)) scattering. As a general observation, it seems that the shear component carries the dominant part of the cloaking error, in the first quadrant of plots (sp30-c) and (ss30-c) most notably. That is because the cloak is based on the behavior of the lattice in the homogenization limit and, at equal frequencies, shear waves propagate at almost half the wavelength of pressure waves. Thus, reducing the cloaking error further requires using a smaller discretization step $\delta$ and a larger numerical effort while the design remains essentially the same.

In conclusion, transformation-based cloaking for mechanical waves has progressed in a number of important cases mostly pertaining to acoustic waves in fluids~\cite{Cummer2007, Chen2007, Norris2008, Norris2015}, to flexural waves in plates~\cite{Farhat2009a, Farhat2009, Farhat2012, Stenger2012, Brun2014, Colquitt2014a, Chen2016d} and to scalar fields more generally~\cite{Parnell2012, Parnell2012a, Buckmann2014, Buckmann2015}. Progress in full elasticity has been slower, impeded by the absence of materials whose behavior is invariant under curvilinear changes of coordinates~\citep{Norris2012a}. Such materials can be designed nonetheless as lattices of subwavelength structural elements~\cite{Nassar2018a, Nassar2019}. In that process, resonance plays a role that goes beyond reflection and refraction and unlocks atypical stress-strain relationships. Here, rotational resonance permitted to break stress and mirror symmetries and granted the elasticity tensor two extra degrees of freedom, polarity and chirality, rich enough to enable cloaking. In this way, the present work established the first theoretical and numerical evidence of resonance-based cloaking in full plane elasticity thanks to the proposed polar metamaterials. Finally, it is worth noting that a major drawback common to all resonance-based solutions, cloaking-related or not, is their dependence over frequency. Overcoming this limitation without involving extrinsic interactions, would constitute a significant step towards applications.
\begin{acknowledgments}
This work is supported by the Army Research Office under Grant No. W911NF-18-1-0031 with Program Manager Dr. David M. Stepp.
\end{acknowledgments}

\begin{thebibliography}{43}%
\makeatletter
\providecommand \@ifxundefined [1]{%
 \@ifx{#1\undefined}
}%
\providecommand \@ifnum [1]{%
 \ifnum #1\expandafter \@firstoftwo
 \else \expandafter \@secondoftwo
 \fi
}%
\providecommand \@ifx [1]{%
 \ifx #1\expandafter \@firstoftwo
 \else \expandafter \@secondoftwo
 \fi
}%
\providecommand \natexlab [1]{#1}%
\providecommand \enquote  [1]{``#1''}%
\providecommand \bibnamefont  [1]{#1}%
\providecommand \bibfnamefont [1]{#1}%
\providecommand \citenamefont [1]{#1}%
\providecommand \href@noop [0]{\@secondoftwo}%
\providecommand \href [0]{\begingroup \@sanitize@url \@href}%
\providecommand \@href[1]{\@@startlink{#1}\@@href}%
\providecommand \@@href[1]{\endgroup#1\@@endlink}%
\providecommand \@sanitize@url [0]{\catcode `\\12\catcode `\$12\catcode
  `\&12\catcode `\#12\catcode `\^12\catcode `\_12\catcode `\%12\relax}%
\providecommand \@@startlink[1]{}%
\providecommand \@@endlink[0]{}%
\providecommand \url  [0]{\begingroup\@sanitize@url \@url }%
\providecommand \@url [1]{\endgroup\@href {#1}{\urlprefix }}%
\providecommand \urlprefix  [0]{URL }%
\providecommand \Eprint [0]{\href }%
\providecommand \doibase [0]{http://dx.doi.org/}%
\providecommand \selectlanguage [0]{\@gobble}%
\providecommand \bibinfo  [0]{\@secondoftwo}%
\providecommand \bibfield  [0]{\@secondoftwo}%
\providecommand \translation [1]{[#1]}%
\providecommand \BibitemOpen [0]{}%
\providecommand \bibitemStop [0]{}%
\providecommand \bibitemNoStop [0]{.\EOS\space}%
\providecommand \EOS [0]{\spacefactor3000\relax}%
\providecommand \BibitemShut  [1]{\csname bibitem#1\endcsname}%
\let\auto@bib@innerbib\@empty
\bibitem [{\citenamefont {Ma}\ and\ \citenamefont {Sheng}(2016)}]{Ma2016}%
  \BibitemOpen
  \bibfield  {author} {\bibinfo {author} {\bibfnamefont {G.}~\bibnamefont
  {Ma}}\ and\ \bibinfo {author} {\bibfnamefont {P.}~\bibnamefont {Sheng}},\
  }\href@noop {} {\bibfield  {journal} {\bibinfo  {journal} {Sci. Adv.}\
  }\textbf {\bibinfo {volume} {2}},\ \bibinfo {pages} {1501595} (\bibinfo
  {year} {2016})}\BibitemShut {NoStop}%
\bibitem [{\citenamefont {Zhou}, \citenamefont {Liu},\ and\ \citenamefont
  {Hu}(2012)}]{Zhou2012}%
  \BibitemOpen
  \bibfield  {author} {\bibinfo {author} {\bibfnamefont {X.}~\bibnamefont
  {Zhou}}, \bibinfo {author} {\bibfnamefont {X.}~\bibnamefont {Liu}}, \ and\
  \bibinfo {author} {\bibfnamefont {G.}~\bibnamefont {Hu}},\ }\href@noop {}
  {\bibfield  {journal} {\bibinfo  {journal} {Theor. Appl. Mech. Lett.}\
  }\textbf {\bibinfo {volume} {2}},\ \bibinfo {pages} {041001} (\bibinfo {year}
  {2012})}\BibitemShut {NoStop}%
\bibitem [{\citenamefont {Liu}\ \emph {et~al.}(2011{\natexlab{a}})\citenamefont
  {Liu}, \citenamefont {Hu}, \citenamefont {Huang},\ and\ \citenamefont
  {Sun}}]{Liu2011}%
  \BibitemOpen
  \bibfield  {author} {\bibinfo {author} {\bibfnamefont {X.~N.}\ \bibnamefont
  {Liu}}, \bibinfo {author} {\bibfnamefont {G.~K.}\ \bibnamefont {Hu}},
  \bibinfo {author} {\bibfnamefont {G.~L.}\ \bibnamefont {Huang}}, \ and\
  \bibinfo {author} {\bibfnamefont {C.~T.}\ \bibnamefont {Sun}},\ }\href@noop
  {} {\bibfield  {journal} {\bibinfo  {journal} {Appl. Phys. Lett.}\ }\textbf
  {\bibinfo {volume} {98}},\ \bibinfo {pages} {1} (\bibinfo {year}
  {2011}{\natexlab{a}})}\BibitemShut {NoStop}%
\bibitem [{\citenamefont {Zhu}\ \emph {et~al.}(2014)\citenamefont {Zhu},
  \citenamefont {Liu}, \citenamefont {Hu}, \citenamefont {Sun},\ and\
  \citenamefont {Huang}}]{Zhu2014}%
  \BibitemOpen
  \bibfield  {author} {\bibinfo {author} {\bibfnamefont {R.}~\bibnamefont
  {Zhu}}, \bibinfo {author} {\bibfnamefont {X.~N.}\ \bibnamefont {Liu}},
  \bibinfo {author} {\bibfnamefont {G.~K.}\ \bibnamefont {Hu}}, \bibinfo
  {author} {\bibfnamefont {C.~T.}\ \bibnamefont {Sun}}, \ and\ \bibinfo
  {author} {\bibfnamefont {G.~L.}\ \bibnamefont {Huang}},\ }\href@noop {}
  {\bibfield  {journal} {\bibinfo  {journal} {Nat. Commun.}\ }\textbf {\bibinfo
  {volume} {5}},\ \bibinfo {pages} {5510} (\bibinfo {year} {2014})}\BibitemShut
  {NoStop}%
\bibitem [{\citenamefont {Landau}\ and\ \citenamefont
  {Lifshitz}(1970)}]{landau1970}%
  \BibitemOpen
  \bibfield  {author} {\bibinfo {author} {\bibfnamefont {L.~D.}\ \bibnamefont
  {Landau}}\ and\ \bibinfo {author} {\bibfnamefont {E.~M.}\ \bibnamefont
  {Lifshitz}},\ }\href@noop {} {\emph {\bibinfo {title} {{Theory of
  Elasticity}}}}\ (\bibinfo  {publisher} {Pergamon Press},\ \bibinfo {address}
  {Oxford},\ \bibinfo {year} {1970})\BibitemShut {NoStop}%
\bibitem [{\citenamefont {Forte}\ and\ \citenamefont
  {Vianello}(1996)}]{Forte1996}%
  \BibitemOpen
  \bibfield  {author} {\bibinfo {author} {\bibfnamefont {S.}~\bibnamefont
  {Forte}}\ and\ \bibinfo {author} {\bibfnamefont {M.}~\bibnamefont
  {Vianello}},\ }\href@noop {} {\bibfield  {journal} {\bibinfo  {journal} {J.
  Elast.}\ }\textbf {\bibinfo {volume} {43}},\ \bibinfo {pages} {81} (\bibinfo
  {year} {1996})}\BibitemShut {NoStop}%
\bibitem [{\citenamefont {He}\ and\ \citenamefont {Zheng}(1996)}]{He1996}%
  \BibitemOpen
  \bibfield  {author} {\bibinfo {author} {\bibfnamefont {Q.-C.}\ \bibnamefont
  {He}}\ and\ \bibinfo {author} {\bibfnamefont {Q.~S.}\ \bibnamefont {Zheng}},\
  }\href@noop {} {\bibfield  {journal} {\bibinfo  {journal} {J. Elast.}\
  }\textbf {\bibinfo {volume} {43}},\ \bibinfo {pages} {203} (\bibinfo {year}
  {1996})}\BibitemShut {NoStop}%
\bibitem [{\citenamefont {{Guevara Vasquez}}\ \emph {et~al.}(2012)\citenamefont
  {{Guevara Vasquez}}, \citenamefont {Milton}, \citenamefont {Onofrei},\ and\
  \citenamefont {Seppecher}}]{GVasquez2012}%
  \BibitemOpen
  \bibfield  {author} {\bibinfo {author} {\bibfnamefont {F.}~\bibnamefont
  {{Guevara Vasquez}}}, \bibinfo {author} {\bibfnamefont {G.~W.}\ \bibnamefont
  {Milton}}, \bibinfo {author} {\bibfnamefont {D.}~\bibnamefont {Onofrei}}, \
  and\ \bibinfo {author} {\bibfnamefont {P.}~\bibnamefont {Seppecher}},\ }in\
  \href@noop {} {\emph {\bibinfo {booktitle} {Acoust. metamaterials Negat.
  Refract. imaging, lensing cloaking}}},\ \bibinfo {series and number}
  {Springer Series in Materials Science},\ \bibinfo {editor} {edited by\
  \bibinfo {editor} {\bibfnamefont {R.~V.}\ \bibnamefont {Craster}}\ and\
  \bibinfo {editor} {\bibfnamefont {S.}~\bibnamefont {Guenneau}}}\ (\bibinfo
  {publisher} {Springer},\ \bibinfo {address} {Dordrecht},\ \bibinfo {year}
  {2012})\ pp.\ \bibinfo {pages} {289--318}\BibitemShut {NoStop}%
\bibitem [{\citenamefont {Nassar}, \citenamefont {Chen},\ and\ \citenamefont
  {Huang}(2019)}]{Nassar2019}%
  \BibitemOpen
  \bibfield  {author} {\bibinfo {author} {\bibfnamefont {H.}~\bibnamefont
  {Nassar}}, \bibinfo {author} {\bibfnamefont {Y.~Y.}\ \bibnamefont {Chen}}, \
  and\ \bibinfo {author} {\bibfnamefont {G.~L.}\ \bibnamefont {Huang}},\
  }\href@noop {} {\bibfield  {journal} {\bibinfo  {journal} {J. Mech. Phys.
  Solids}\ }\textbf {\bibinfo {volume} {129}},\ \bibinfo {pages} {229}
  (\bibinfo {year} {2019})}\BibitemShut {NoStop}%
\bibitem [{\citenamefont {Nassar}, \citenamefont {Chen},\ and\ \citenamefont
  {Huang}(2018)}]{Nassar2018a}%
  \BibitemOpen
  \bibfield  {author} {\bibinfo {author} {\bibfnamefont {H.}~\bibnamefont
  {Nassar}}, \bibinfo {author} {\bibfnamefont {Y.}~\bibnamefont {Chen}}, \ and\
  \bibinfo {author} {\bibfnamefont {G.~L.}\ \bibnamefont {Huang}},\ }\href@noop
  {} {\bibfield  {journal} {\bibinfo  {journal} {Proc. R. Soc. A}\ }\textbf
  {\bibinfo {volume} {474}},\ \bibinfo {pages} {20180523} (\bibinfo {year}
  {2018})}\BibitemShut {NoStop}%
\bibitem [{\citenamefont {Liu}, \citenamefont {Chan},\ and\ \citenamefont
  {Sheng}(2005)}]{Liu2005}%
  \BibitemOpen
  \bibfield  {author} {\bibinfo {author} {\bibfnamefont {Z.}~\bibnamefont
  {Liu}}, \bibinfo {author} {\bibfnamefont {C.~T.}\ \bibnamefont {Chan}}, \
  and\ \bibinfo {author} {\bibfnamefont {P.}~\bibnamefont {Sheng}},\
  }\href@noop {} {\bibfield  {journal} {\bibinfo  {journal} {Phys. Rev. B}\
  }\textbf {\bibinfo {volume} {71}},\ \bibinfo {pages} {1} (\bibinfo {year}
  {2005})}\BibitemShut {NoStop}%
\bibitem [{\citenamefont {Favier}\ \emph {et~al.}(2018)\citenamefont {Favier},
  \citenamefont {Nemati}, \citenamefont {Perrot},\ and\ \citenamefont
  {He}}]{Favier2018}%
  \BibitemOpen
  \bibfield  {author} {\bibinfo {author} {\bibfnamefont {E.}~\bibnamefont
  {Favier}}, \bibinfo {author} {\bibfnamefont {N.}~\bibnamefont {Nemati}},
  \bibinfo {author} {\bibfnamefont {C.}~\bibnamefont {Perrot}}, \ and\ \bibinfo
  {author} {\bibfnamefont {Q.-C.}\ \bibnamefont {He}},\ }\href@noop {}
  {\bibfield  {journal} {\bibinfo  {journal} {J. Phys. Commun.}\ }\textbf
  {\bibinfo {volume} {2}},\ \bibinfo {pages} {035035} (\bibinfo {year}
  {2018})}\BibitemShut {NoStop}%
\bibitem [{Note1()}]{Note1}%
  \BibitemOpen
  \bibinfo {note} {See Supplemental Material for further detail.}\BibitemShut
  {Stop}%
\bibitem [{\citenamefont {Liu}\ \emph {et~al.}(2011{\natexlab{b}})\citenamefont
  {Liu}, \citenamefont {Hu}, \citenamefont {Sun},\ and\ \citenamefont
  {Huang}}]{Liu2011a}%
  \BibitemOpen
  \bibfield  {author} {\bibinfo {author} {\bibfnamefont {X.~N.}\ \bibnamefont
  {Liu}}, \bibinfo {author} {\bibfnamefont {G.~K.}\ \bibnamefont {Hu}},
  \bibinfo {author} {\bibfnamefont {C.~T.}\ \bibnamefont {Sun}}, \ and\
  \bibinfo {author} {\bibfnamefont {G.~L.}\ \bibnamefont {Huang}},\ }\href@noop
  {} {\bibfield  {journal} {\bibinfo  {journal} {J. Sound Vib.}\ }\textbf
  {\bibinfo {volume} {330}},\ \bibinfo {pages} {2536} (\bibinfo {year}
  {2011}{\natexlab{b}})}\BibitemShut {NoStop}%
\bibitem [{\citenamefont {Spadoni}\ and\ \citenamefont
  {Ruzzene}(2012)}]{Spadoni2012}%
  \BibitemOpen
  \bibfield  {author} {\bibinfo {author} {\bibfnamefont {A.}~\bibnamefont
  {Spadoni}}\ and\ \bibinfo {author} {\bibfnamefont {M.}~\bibnamefont
  {Ruzzene}},\ }\href@noop {} {\bibfield  {journal} {\bibinfo  {journal} {J.
  Mech. Phys. Solids}\ }\textbf {\bibinfo {volume} {60}},\ \bibinfo {pages}
  {156} (\bibinfo {year} {2012})}\BibitemShut {NoStop}%
\bibitem [{\citenamefont {Liu}, \citenamefont {Huang},\ and\ \citenamefont
  {Hu}(2012)}]{Liu2012a}%
  \BibitemOpen
  \bibfield  {author} {\bibinfo {author} {\bibfnamefont {X.~N.}\ \bibnamefont
  {Liu}}, \bibinfo {author} {\bibfnamefont {G.~L.}\ \bibnamefont {Huang}}, \
  and\ \bibinfo {author} {\bibfnamefont {G.~K.}\ \bibnamefont {Hu}},\
  }\href@noop {} {\bibfield  {journal} {\bibinfo  {journal} {J. Mech. Phys.
  Solids}\ }\textbf {\bibinfo {volume} {60}},\ \bibinfo {pages} {1907}
  (\bibinfo {year} {2012})}\BibitemShut {NoStop}%
\bibitem [{\citenamefont {Chen}, \citenamefont {Liu},\ and\ \citenamefont
  {Hu}(2014)}]{Chen2014b}%
  \BibitemOpen
  \bibfield  {author} {\bibinfo {author} {\bibfnamefont {Y.}~\bibnamefont
  {Chen}}, \bibinfo {author} {\bibfnamefont {X.}~\bibnamefont {Liu}}, \ and\
  \bibinfo {author} {\bibfnamefont {G.}~\bibnamefont {Hu}},\ }\href@noop {}
  {\bibfield  {journal} {\bibinfo  {journal} {Comptes Rendus - Mec.}\ }\textbf
  {\bibinfo {volume} {342}},\ \bibinfo {pages} {273} (\bibinfo {year}
  {2014})}\BibitemShut {NoStop}%
\bibitem [{\citenamefont {Frenzel}, \citenamefont {Kadic},\ and\ \citenamefont
  {Wegener}(2017)}]{Frenzel2017}%
  \BibitemOpen
  \bibfield  {author} {\bibinfo {author} {\bibfnamefont {T.}~\bibnamefont
  {Frenzel}}, \bibinfo {author} {\bibfnamefont {M.}~\bibnamefont {Kadic}}, \
  and\ \bibinfo {author} {\bibfnamefont {M.}~\bibnamefont {Wegener}},\
  }\href@noop {} {\bibfield  {journal} {\bibinfo  {journal} {Science}\
  }\textbf {\bibinfo {volume} {358}},\ \bibinfo {pages} {1072} (\bibinfo {year}
  {2017})}\BibitemShut {NoStop}%
\bibitem [{\citenamefont {Auffray}, \citenamefont {Bouchet},\ and\
  \citenamefont {Br{\'{e}}chet}(2010)}]{Auffray2010}%
  \BibitemOpen
  \bibfield  {author} {\bibinfo {author} {\bibfnamefont {N.}~\bibnamefont
  {Auffray}}, \bibinfo {author} {\bibfnamefont {R.}~\bibnamefont {Bouchet}}, \
  and\ \bibinfo {author} {\bibfnamefont {Y.}~\bibnamefont {Br{\'{e}}chet}},\
  }\href@noop {} {\bibfield  {journal} {\bibinfo  {journal} {Int. J. Solids
  Struct.}\ }\textbf {\bibinfo {volume} {47}},\ \bibinfo {pages} {1698}
  (\bibinfo {year} {2010})}\BibitemShut {NoStop}%
\bibitem [{\citenamefont {Bacigalupo}\ and\ \citenamefont
  {Gambarotta}(2014)}]{Bacigalupo2014}%
  \BibitemOpen
  \bibfield  {author} {\bibinfo {author} {\bibfnamefont {A.}~\bibnamefont
  {Bacigalupo}}\ and\ \bibinfo {author} {\bibfnamefont {L.}~\bibnamefont
  {Gambarotta}},\ }\href@noop {} {\bibfield  {journal} {\bibinfo  {journal}
  {Compos. Struct.}\ }\textbf {\bibinfo {volume} {116}},\ \bibinfo {pages}
  {461} (\bibinfo {year} {2014})}\BibitemShut {NoStop}%
\bibitem [{\citenamefont {Rosi}\ and\ \citenamefont
  {Auffray}(2016)}]{Rosi2016}%
  \BibitemOpen
  \bibfield  {author} {\bibinfo {author} {\bibfnamefont {G.}~\bibnamefont
  {Rosi}}\ and\ \bibinfo {author} {\bibfnamefont {N.}~\bibnamefont {Auffray}},\
  }\href@noop {} {\bibfield  {journal} {\bibinfo  {journal} {Wave Motion}\
  }\textbf {\bibinfo {volume} {63}},\ \bibinfo {pages} {120} (\bibinfo {year}
  {2016})}\BibitemShut {NoStop}%
\bibitem [{\citenamefont {Norris}, \citenamefont {Shuvalov},\ and\
  \citenamefont {Kutsenko}(2012)}]{Norris2012}%
  \BibitemOpen
  \bibfield  {author} {\bibinfo {author} {\bibfnamefont {A.~N.}\ \bibnamefont
  {Norris}}, \bibinfo {author} {\bibfnamefont {A.~L.}\ \bibnamefont
  {Shuvalov}}, \ and\ \bibinfo {author} {\bibfnamefont {A.~A.}\ \bibnamefont
  {Kutsenko}},\ }\href@noop {} {\bibfield  {journal} {\bibinfo  {journal}
  {Proc. R. Soc. A}\ }\textbf {\bibinfo {volume} {468}},\ \bibinfo {pages}
  {1629} (\bibinfo {year} {2012})}\BibitemShut {NoStop}%
\bibitem [{\citenamefont {Xu}\ and\ \citenamefont {Chen}(2014)}]{Xu2014}%
  \BibitemOpen
  \bibfield  {author} {\bibinfo {author} {\bibfnamefont {L.}~\bibnamefont
  {Xu}}\ and\ \bibinfo {author} {\bibfnamefont {H.}~\bibnamefont {Chen}},\
  }\href@noop {} {\bibfield  {journal} {\bibinfo  {journal} {Nat. Photonics}\
  }\textbf {\bibinfo {volume} {9}},\ \bibinfo {pages} {15} (\bibinfo {year}
  {2014})}\BibitemShut {NoStop}%
\bibitem [{\citenamefont {Greenleaf}, \citenamefont {Lassas},\ and\
  \citenamefont {Uhlmann}(2003{\natexlab{a}})}]{Greenleaf2003a}%
  \BibitemOpen
  \bibfield  {author} {\bibinfo {author} {\bibfnamefont {A.}~\bibnamefont
  {Greenleaf}}, \bibinfo {author} {\bibfnamefont {M.}~\bibnamefont {Lassas}}, \
  and\ \bibinfo {author} {\bibfnamefont {G.}~\bibnamefont {Uhlmann}},\
  }\href@noop {} {\bibfield  {journal} {\bibinfo  {journal} {Physiol. Meas.}\
  }\textbf {\bibinfo {volume} {24}},\ \bibinfo {pages} {413} (\bibinfo {year}
  {2003}{\natexlab{a}})}\BibitemShut {NoStop}%
\bibitem [{\citenamefont {Greenleaf}, \citenamefont {Lassas},\ and\
  \citenamefont {Uhlmann}(2003{\natexlab{b}})}]{Greenleaf2003}%
  \BibitemOpen
  \bibfield  {author} {\bibinfo {author} {\bibfnamefont {A.}~\bibnamefont
  {Greenleaf}}, \bibinfo {author} {\bibfnamefont {M.}~\bibnamefont {Lassas}}, \
  and\ \bibinfo {author} {\bibfnamefont {G.}~\bibnamefont {Uhlmann}},\
  }\href@noop {} {\bibfield  {journal} {\bibinfo  {journal} {Math. Res. Lett.}\
  }\textbf {\bibinfo {volume} {10}},\ \bibinfo {pages} {685} (\bibinfo {year}
  {2003}{\natexlab{b}})}\BibitemShut {NoStop}%
\bibitem [{\citenamefont {Leonhardt}(2006)}]{Leonhardt2006}%
  \BibitemOpen
  \bibfield  {author} {\bibinfo {author} {\bibfnamefont {U.}~\bibnamefont
  {Leonhardt}},\ }\href@noop {} {\bibfield  {journal} {\bibinfo  {journal}
  {Science}\ }\textbf {\bibinfo {volume} {312}},\ \bibinfo {pages} {1777}
  (\bibinfo {year} {2006})}\BibitemShut {NoStop}%
\bibitem [{\citenamefont {Pendry}, \citenamefont {Schurig},\ and\ \citenamefont
  {Smith}(2006)}]{Pendry2006}%
  \BibitemOpen
  \bibfield  {author} {\bibinfo {author} {\bibfnamefont {J.~B.}\ \bibnamefont
  {Pendry}}, \bibinfo {author} {\bibfnamefont {D.}~\bibnamefont {Schurig}}, \
  and\ \bibinfo {author} {\bibfnamefont {D.~R.}\ \bibnamefont {Smith}},\
  }\href@noop {} {\bibfield  {journal} {\bibinfo  {journal} {Science}\ }\textbf
  {\bibinfo {volume} {312}},\ \bibinfo {pages} {1780} (\bibinfo {year}
  {2006})}\BibitemShut {NoStop}%
\bibitem [{\citenamefont {Cummer}\ and\ \citenamefont
  {Schurig}(2007)}]{Cummer2007}%
  \BibitemOpen
  \bibfield  {author} {\bibinfo {author} {\bibfnamefont {S.~A.}\ \bibnamefont
  {Cummer}}\ and\ \bibinfo {author} {\bibfnamefont {D.}~\bibnamefont
  {Schurig}},\ }\href@noop {} {\bibfield  {journal} {\bibinfo  {journal} {New
  J. Phys.}\ }\textbf {\bibinfo {volume} {9}},\ \bibinfo {pages} {45} (\bibinfo
  {year} {2007})}\BibitemShut {NoStop}%
\bibitem [{\citenamefont {Chen}\ and\ \citenamefont {Chan}(2007)}]{Chen2007}%
  \BibitemOpen
  \bibfield  {author} {\bibinfo {author} {\bibfnamefont {H.}~\bibnamefont
  {Chen}}\ and\ \bibinfo {author} {\bibfnamefont {C.~T.}\ \bibnamefont
  {Chan}},\ }\href@noop {} {\bibfield  {journal} {\bibinfo  {journal} {Appl.
  Phys. Lett.}\ }\textbf {\bibinfo {volume} {91}},\ \bibinfo {pages} {183518}
  (\bibinfo {year} {2007})}\BibitemShut {NoStop}%
\bibitem [{\citenamefont {Norris}(2008)}]{Norris2008}%
  \BibitemOpen
  \bibfield  {author} {\bibinfo {author} {\bibfnamefont {A.~N.}\ \bibnamefont
  {Norris}},\ }\href@noop {} {\bibfield  {journal} {\bibinfo  {journal} {Proc.
  R. Soc. A}\ }\textbf {\bibinfo {volume} {464}},\ \bibinfo {pages} {2411}
  (\bibinfo {year} {2008})}\BibitemShut {NoStop}%
\bibitem [{\citenamefont {Norris}(2015)}]{Norris2015}%
  \BibitemOpen
  \bibfield  {author} {\bibinfo {author} {\bibfnamefont {A.~N.}\ \bibnamefont
  {Norris}},\ }\href@noop {} {\bibfield  {journal} {\bibinfo  {journal}
  {Acoust. Today}\ }\textbf {\bibinfo {volume} {11}},\ \bibinfo {pages} {38}
  (\bibinfo {year} {2015})}\BibitemShut {NoStop}%
\bibitem [{\citenamefont {Farhat}, \citenamefont {Guenneau},\ and\
  \citenamefont {Enoch}(2009)}]{Farhat2009a}%
  \BibitemOpen
  \bibfield  {author} {\bibinfo {author} {\bibfnamefont {M.}~\bibnamefont
  {Farhat}}, \bibinfo {author} {\bibfnamefont {S.}~\bibnamefont {Guenneau}}, \
  and\ \bibinfo {author} {\bibfnamefont {S.}~\bibnamefont {Enoch}},\
  }\href@noop {} {\bibfield  {journal} {\bibinfo  {journal} {Phys. Rev. Lett.}\
  }\textbf {\bibinfo {volume} {103}},\ \bibinfo {pages} {024301} (\bibinfo
  {year} {2009})}\BibitemShut {NoStop}%
\bibitem [{\citenamefont {Farhat}\ \emph {et~al.}(2009)\citenamefont {Farhat},
  \citenamefont {Guenneau}, \citenamefont {Enoch},\ and\ \citenamefont
  {Movchan}}]{Farhat2009}%
  \BibitemOpen
  \bibfield  {author} {\bibinfo {author} {\bibfnamefont {M.}~\bibnamefont
  {Farhat}}, \bibinfo {author} {\bibfnamefont {S.}~\bibnamefont {Guenneau}},
  \bibinfo {author} {\bibfnamefont {S.}~\bibnamefont {Enoch}}, \ and\ \bibinfo
  {author} {\bibfnamefont {A.~B.}\ \bibnamefont {Movchan}},\ }\href@noop {}
  {\bibfield  {journal} {\bibinfo  {journal} {Phys. Rev. B - Condens. Matter
  Mater. Phys.}\ }\textbf {\bibinfo {volume} {79}},\ \bibinfo {pages} {033102}
  (\bibinfo {year} {2009})}\BibitemShut {NoStop}%
\bibitem [{\citenamefont {Farhat}, \citenamefont {Guenneau},\ and\
  \citenamefont {Enoch}(2012)}]{Farhat2012}%
  \BibitemOpen
  \bibfield  {author} {\bibinfo {author} {\bibfnamefont {M.}~\bibnamefont
  {Farhat}}, \bibinfo {author} {\bibfnamefont {S.}~\bibnamefont {Guenneau}}, \
  and\ \bibinfo {author} {\bibfnamefont {S.}~\bibnamefont {Enoch}},\
  }\href@noop {} {\bibfield  {journal} {\bibinfo  {journal} {Phys. Rev. B -
  Condens. Matter Mater. Phys.}\ }\textbf {\bibinfo {volume} {85}},\ \bibinfo
  {pages} {020301} (\bibinfo {year} {2012})}\BibitemShut {NoStop}%
\bibitem [{\citenamefont {Stenger}, \citenamefont {Wilhelm},\ and\
  \citenamefont {Wegener}(2012)}]{Stenger2012}%
  \BibitemOpen
  \bibfield  {author} {\bibinfo {author} {\bibfnamefont {N.}~\bibnamefont
  {Stenger}}, \bibinfo {author} {\bibfnamefont {M.}~\bibnamefont {Wilhelm}}, \
  and\ \bibinfo {author} {\bibfnamefont {M.}~\bibnamefont {Wegener}},\
  }\href@noop {} {\bibfield  {journal} {\bibinfo  {journal} {Phys. Rev. Lett.}\
  }\textbf {\bibinfo {volume} {108}},\ \bibinfo {pages} {014301} (\bibinfo
  {year} {2012})}\BibitemShut {NoStop}%
\bibitem [{\citenamefont {Brun}\ \emph {et~al.}(2014)\citenamefont {Brun},
  \citenamefont {Colquitt}, \citenamefont {Jones}, \citenamefont {Movchan},\
  and\ \citenamefont {Movchan}}]{Brun2014}%
  \BibitemOpen
  \bibfield  {author} {\bibinfo {author} {\bibfnamefont {M.}~\bibnamefont
  {Brun}}, \bibinfo {author} {\bibfnamefont {D.~J.}\ \bibnamefont {Colquitt}},
  \bibinfo {author} {\bibfnamefont {I.~S.}\ \bibnamefont {Jones}}, \bibinfo
  {author} {\bibfnamefont {A.~B.}\ \bibnamefont {Movchan}}, \ and\ \bibinfo
  {author} {\bibfnamefont {N.~V.}\ \bibnamefont {Movchan}},\ }\href@noop {}
  {\bibfield  {journal} {\bibinfo  {journal} {New J. Phys.}\ }\textbf {\bibinfo
  {volume} {16}},\ \bibinfo {pages} {093020} (\bibinfo {year}
  {2014})}\BibitemShut {NoStop}%
\bibitem [{\citenamefont {Colquitt}\ \emph {et~al.}(2014)\citenamefont
  {Colquitt}, \citenamefont {Brun}, \citenamefont {Gei}, \citenamefont
  {Movchan}, \citenamefont {Movchan},\ and\ \citenamefont
  {Jones}}]{Colquitt2014a}%
  \BibitemOpen
  \bibfield  {author} {\bibinfo {author} {\bibfnamefont {D.~J.}\ \bibnamefont
  {Colquitt}}, \bibinfo {author} {\bibfnamefont {M.}~\bibnamefont {Brun}},
  \bibinfo {author} {\bibfnamefont {M.}~\bibnamefont {Gei}}, \bibinfo {author}
  {\bibfnamefont {A.~B.}\ \bibnamefont {Movchan}}, \bibinfo {author}
  {\bibfnamefont {N.~V.}\ \bibnamefont {Movchan}}, \ and\ \bibinfo {author}
  {\bibfnamefont {I.~S.}\ \bibnamefont {Jones}},\ }\href@noop {} {\bibfield
  {journal} {\bibinfo  {journal} {J. Mech. Phys. Solids}\ }\textbf {\bibinfo
  {volume} {72}},\ \bibinfo {pages} {131} (\bibinfo {year} {2014})}\BibitemShut
  {NoStop}%
\bibitem [{\citenamefont {Chen}, \citenamefont {Hu},\ and\ \citenamefont
  {Huang}(2016)}]{Chen2016d}%
  \BibitemOpen
  \bibfield  {author} {\bibinfo {author} {\bibfnamefont {Y.}~\bibnamefont
  {Chen}}, \bibinfo {author} {\bibfnamefont {J.}~\bibnamefont {Hu}}, \ and\
  \bibinfo {author} {\bibfnamefont {G.~L.}\ \bibnamefont {Huang}},\ }\href@noop
  {} {\bibfield  {journal} {\bibinfo  {journal} {J. Intell. Mater. Syst.
  Struct.}\ }\textbf {\bibinfo {volume} {27}},\ \bibinfo {pages} {1337}
  (\bibinfo {year} {2016})}\BibitemShut {NoStop}%
\bibitem [{\citenamefont {Parnell}, \citenamefont {Norris},\ and\ \citenamefont
  {Shearer}(2012)}]{Parnell2012}%
  \BibitemOpen
  \bibfield  {author} {\bibinfo {author} {\bibfnamefont {W.~J.}\ \bibnamefont
  {Parnell}}, \bibinfo {author} {\bibfnamefont {A.~N.}\ \bibnamefont {Norris}},
  \ and\ \bibinfo {author} {\bibfnamefont {T.}~\bibnamefont {Shearer}},\
  }\href@noop {} {\bibfield  {journal} {\bibinfo  {journal} {Appl. Phys.
  Lett.}\ }\textbf {\bibinfo {volume} {100}},\ \bibinfo {pages} {171907}
  (\bibinfo {year} {2012})}\BibitemShut {NoStop}%
\bibitem [{\citenamefont {Parnell}(2012)}]{Parnell2012a}%
  \BibitemOpen
  \bibfield  {author} {\bibinfo {author} {\bibfnamefont {W.~J.}\ \bibnamefont
  {Parnell}},\ }\href@noop {} {\bibfield  {journal} {\bibinfo  {journal} {Proc.
  R. Soc. A}\ }\textbf {\bibinfo {volume} {468}},\ \bibinfo {pages} {563}
  (\bibinfo {year} {2012})}\BibitemShut {NoStop}%
\bibitem [{\citenamefont {B{\"{u}}ckmann}\ \emph {et~al.}(2014)\citenamefont
  {B{\"{u}}ckmann}, \citenamefont {Thiel}, \citenamefont {Kadic}, \citenamefont
  {Schittny},\ and\ \citenamefont {Wegener}}]{Buckmann2014}%
  \BibitemOpen
  \bibfield  {author} {\bibinfo {author} {\bibfnamefont {T.}~\bibnamefont
  {B{\"{u}}ckmann}}, \bibinfo {author} {\bibfnamefont {M.}~\bibnamefont
  {Thiel}}, \bibinfo {author} {\bibfnamefont {M.}~\bibnamefont {Kadic}},
  \bibinfo {author} {\bibfnamefont {R.}~\bibnamefont {Schittny}}, \ and\
  \bibinfo {author} {\bibfnamefont {M.}~\bibnamefont {Wegener}},\ }\href@noop
  {} {\bibfield  {journal} {\bibinfo  {journal} {Nat. Commun.}\ }\textbf
  {\bibinfo {volume} {5}},\ \bibinfo {pages} {4130} (\bibinfo {year}
  {2014})}\BibitemShut {NoStop}%
\bibitem [{\citenamefont {B{\"{u}}ckmann}\ \emph {et~al.}(2015)\citenamefont
  {B{\"{u}}ckmann}, \citenamefont {Kadic}, \citenamefont {Schittny},\ and\
  \citenamefont {Wegener}}]{Buckmann2015}%
  \BibitemOpen
  \bibfield  {author} {\bibinfo {author} {\bibfnamefont {T.}~\bibnamefont
  {B{\"{u}}ckmann}}, \bibinfo {author} {\bibfnamefont {M.}~\bibnamefont
  {Kadic}}, \bibinfo {author} {\bibfnamefont {R.}~\bibnamefont {Schittny}}, \
  and\ \bibinfo {author} {\bibfnamefont {M.}~\bibnamefont {Wegener}},\
  }\href@noop {} {\bibfield  {journal} {\bibinfo  {journal} {Proc. Natl. Acad.
  Sci.}\ }\textbf {\bibinfo {volume} {112}},\ \bibinfo {pages} {4930} (\bibinfo
  {year} {2015})}\BibitemShut {NoStop}%
\bibitem [{\citenamefont {Norris}\ and\ \citenamefont
  {Parnell}(2012)}]{Norris2012a}%
  \BibitemOpen
  \bibfield  {author} {\bibinfo {author} {\bibfnamefont {A.~N.}\ \bibnamefont
  {Norris}}\ and\ \bibinfo {author} {\bibfnamefont {W.~J.}\ \bibnamefont
  {Parnell}},\ }\href@noop {} {\bibfield  {journal} {\bibinfo  {journal} {Proc.
  R. Soc. A}\ }\textbf {\bibinfo {volume} {468}},\ \bibinfo {pages} {2881}
  (\bibinfo {year} {2012})}\BibitemShut {NoStop}%
\end{thebibliography}
%
\end{document}